\documentstyle[11pt]{article}

\renewcommand{\thesection}{\Roman{section}}

\def\beq{\vspace{0.2cm} \begin{equation}}
\def\eeq{\vspace{0.2cm} \end{equation} }
\def\beqn{\vspace{0.2cm} \begin{eqnarray}}
\def\eeqn{\vspace{0.2cm} \end{eqnarray} }

\def\bnabla{\mbox{\boldmath $\nabla$}}

\def\ltsim{\raisebox{-.6ex}{$\  \stackrel{\textstyle <}{\textstyle
\sim}\  $}}

\begin{document}

\pagestyle{plain}

\title{\bf {MONTE CARLO ANALYSIS OF A NEW INTERATOMIC POTENTIAL FOR He}}

\author {J. Boronat and J. Casulleras\\ {\it Departament de F\'{\i}sica
i Enginyeria Nuclear,}
\\{\it Universitat Polit\`{e}cnica de Catalunya, Pau Gargallo 5,
E-08028 Barcelona, Spain} }

\date{}

\maketitle

\vspace{0.5cm}

\begin{abstract}
By means of a Quadratic Diffusion Monte Carlo method we have performed a
comparative analysis between the Aziz potential and a revised version
of it.
The results demonstrate that the new potential produces a better
description of
the equation of state for liquid $^4$He. In spite of the improvement in
the
description of derivative magnitudes of the energy, as the pressure or
the compressibility, the energy per particle which comes from this new
potential is lower than the experimental one. The inclusion of
three-body interactions, which give a repulsive contribution to the
potential energy, makes it feasible that the calculated energy
comes close to the experimental result.

\vspace{1.5cm}

\noindent {\normalsize PACS number: 67}

\end{abstract}

\thispagestyle{empty}
\pagebreak

\setcounter{page}{1}
\section{Introduction}

Many-body techniques have achieved a high level of accuracy in the
description of atomic $^3$He and $^4$He, which constitute the
most characteristic examples of quantum liquids.
The theoretical approaches to the many-body problem
can be classified in two large blocks depending on the use or non-use
of stochastic procedures. Among the non-stochastic methods it is the
variational framework \cite{RV88} combined with integral equations
such as HNC, which
has provided the best results in the knowledge of the ground state.
Also, perturbation schemes constructed on correlated basis (correlated
basis function theory -CBF- \cite{FF88}) has led one to extend this
study to
the lowest excited states. On the other hand, Monte Carlo (MC) methods
\cite{Mon91}
give exact information, within some statistical uncertainities, on the
ground state of bosonic systems both at zero and finite temperature.
The initial
constraint imposed by the use of a finite number of particles in MC
simulations does not influence appreciably to the energetic properties.
However, the structure properties at $r \rightarrow
\infty$ ($k \rightarrow 0$) related to long-range correlations are out
of scope.

The high agreement between the theoretical results and the
experimental
data is also linked to the well known interatomic interaction
for He atoms (pair wise additive form). For the last
ten years, the
HFDHE2 potential proposed by Aziz {\em et al.} \cite{Az79} has allowed
for
reproducing the energetic and structure properties of liquid
He quite well both in homogeneous \cite{Ka81} and inhomogeneous phases
\cite{Pa83,Chi92,Va88}. Despite of the accuracy of
this pair-potential a renewed version of it (HFD-B(HE)) was published by
Aziz {\em et al.} in 1987 \cite{Az87}. The new Aziz potential (hereafter
referred to as
Aziz II potential) was brought about as a consequence of several
new theoretical and
experimental results which appeared in the literature between the
publication
of the two potentials. First, Ceperley {\em et al.} \cite{Ce86} pointed
out by means
of a quantum Monte Carlo calculation of the interaction energy of two He
atoms, with internuclear separations less than 1.8 \AA, that the Aziz
potential is too repulsive below this distance. On the other hand, new
experimental measurements of the second virial coefficients and
transport
properties for $^3$He and $^4$He showed evidence of some small
inconsistences of the Aziz potential.
The explicit expressions of the Aziz and Aziz II potentials appear in
the Appendix A. Apart from a soft core, the Aziz II potential has its
minimum at $\varepsilon=10.95\ K$, $r_m=2.963$ \AA\ while Aziz potential
has its minimum at $\varepsilon=10.80\ K$, $r_m=2.967$ \AA. Therefore,
the
new potential is only slightly deeper with the minimum localized at a
lower interatomic separation.

To start on a theoretical comparative study between He potentials it is
necessary to calculate the properties of the liquid as precisely
as it is possible.
Stochastic methods provide the appropriate tools
for this purpose, especially in the case of bosonic systems as
$^4$He. In the past, the Green's function Monte Carlo method (GFMC)
was used to elucidate between different models for the pair interaction.
The main conclusion of this analysis \cite{Ka81} stated that the Aziz
potential was
the best interaction to study the properties of liquid and solid helium.

Our objective in the present work is to perform a comparative analysis
between the two Aziz potentials to establish if the new potential (Aziz
II) produces even better results than the previous one. The calculation
presented here follows an alternative procedure to GFMC known as
Diffusion Monte Carlo (DMC).

Both GFMC, developed by Kalos {\em et al.} \cite{Ce79,Ka74}, and
DMC algorithms \cite{Gu88,Re82} solve
stochastically the Schr\"{o}dinger equation in imaginary time. The GFMC
scheme constructs a time integrated Green's function by means of a
double
Monte Carlo sampling. On the other hand, the DMC algorithm is a simpler
method that assumes an approximate form for the Green's function for
small time-steps $\Delta t$. In this case, after an iterative
process and sufficiently long times, only the ground state wave function
survives. Therefore, the exact energy per particle of the system is
obtained when the limit $\Delta t \rightarrow 0$ is considered. DMC is
posterior to GFMC but up to now it has already been applied
to the study
of small molecules \cite{Re82}, solid hydrogen \cite{Cep87} or $^4$He
clusters \cite{Chi92}. The main
disadvantage of the DMC algorithms used in the major part of those
works
is that the energy eigenvalues change linearly with $\Delta t$. This
fact obliges one to perform several calculations using different
values
for the time-step and next to extrapolate the exact value in the limit
$\Delta t \rightarrow 0$. To avoid this difficulty several quadratic
algorithms have been devised but the success of this improvement
has not
been complete. Recently, a new Quadratic
Diffusion Monte Carlo (QDMC) method \cite{Chi90} has proved to work
efficiently in
the description of $^4$He droplets \cite{Chi92}. In the present work, we
use a QDMC method with a very similar algorithm to the one reported in
Ref.\cite{Chi90}.
In the next sections of the article we will justify the accuracy
of the proposed method which allows for the possibility of calculating
the properties of the system at a finite time-step
without introducing any significant difference with the extrapolated
value.

The outline of this paper is as follows: In Sec. II the Quadratic
Diffusion Monte Carlo method to solve the Schr\"{o}dinger equation is
presented. The consistency of the algorithm is checked by using
different trial functions and several numbers of
particles. The time-step dependence of the energy per particle shows the
expected quadratic behaviour. A comparative analysis of the two Aziz
potentials is reported in Sec. III.
A perturbative estimation of the contributions coming from various
three-body potentials is also reported. A brief discussion and
conclusions comprise Sec. IV.


\section{Computational algorithm}

The starting point in Diffusion Monte Carlo methods is the
Schr\"{o}dinger equation for $N$ particles written in imaginary time:
\beq
- \frac{\partial \Psi({\bf R},t)}{\partial t} = (H-E)\, \Psi({\bf R},t)
\label{eq1}
\eeq
where ${\bf R} \equiv ({\bf r}_1,\ldots,{\bf r}_N)$ and $t$ is measured in
units of $\hbar$. $\Psi({\bf R},t)$ can be expanded in terms of a
complete set of eigenfunctions $\phi_i({\bf R})$ of the Hamiltonian:
\beq
\Psi({\bf R},t)=\sum_{n}c_n \, \exp \left[\, -(E_i-E)t \, \right]\,
\phi_i({\bf R})\ ,
\label{eq1b}
\eeq
where $E_i$ is the eigenvalue associated to $\phi_i({\bf R})$.
The asymptotic
solution of Eq. (\ref{eq1}) for any value $E$ close to the energy
of the ground state and for long times ($t \rightarrow \infty$) gives
$\phi_0({\bf R})$, provided that there is a non-zero overlap between
$\Psi({\bf R},t=0)$ and the ground state wave function $\phi_0({\bf
R})$.

In a computer simulation of Eq. (\ref{eq1}) it is crucial to use the
importance sampling technique \cite{Ka74} in order to reduce the
statistical fluctuations to a manageable level.  Following this
method, one rewrites the Schr\"{o}dinger equation for the function:
\beq
f({\bf R},t)\equiv \psi({\bf R})\,\Psi({\bf R},t)\ ,
\label{eq2}
\eeq
where $\psi({\bf R})$ is a time-independent trial function. Considering a
Hamiltonian of the form:
\beq
H=-\frac{\hbar^2}{2\,m} \, \bnabla^2_{{\bf R}} + V({\bf R})\ ,
\label{eq3}
\eeq
Eq. (\ref{eq1}) turns out to be:
\beqn
-\frac{\partial f({\bf R},t)}{\partial t} & = & -D\, \bnabla^2_{{\bf R}}
f({\bf R},t)+D\, \bnabla_{{\bf R}} \left( {\bf F}({\bf R})\,f({\bf R},t)\,
\right)+\left(E_L({\bf R})-E \right)\,f({\bf R},t) \nonumber  \\[0.6cm]
& \equiv & (A_1+A_2+A_3)\, f({\bf R},t) \equiv A\, f({\bf R},t)\ ,
\label{eq4}
\eeqn
where $D=\hbar^2 /(2m)$, $E_L({\bf R})=\psi({\bf R})^{-1} H \psi({\bf R})$
is the local energy, and
\beq
{\bf F}({\bf R}) = 2\, \psi({\bf R})^{-1} \bnabla_{{\bf R}} \psi({\bf R})
\label{eq5}
\eeq
is called the drift force. ${\bf F}({\bf R})$ acts as an external force
which guides the diffusion process, involved by the first term in Eq.
(\ref{eq4}), to regions where $\psi({\bf R})$ is large.

The formal solution of Eq. (\ref{eq4}) is
\beq
     f({\bf R}^{\prime},t+\Delta t) =\int G({\bf R}^{\prime},{\bf R},
\Delta t)\, f({\bf R},t)\, d{\bf R}
\label{eq6}
\eeq
 with
\beq
    G({\bf R}^{\prime},{\bf R}, \Delta t) = \left \langle\,
{\bf R}^{\prime}\, | \, \exp(-A \Delta t)\, |\, {\bf R}\, \right \rangle.
\label{eq7}
\eeq

While GFMC method works with the whole Green's function,
DMC algorithms rely on reasonable approximations of
$G({\bf R}^{\prime},{\bf R},\Delta t)$ for small values of the time-step
$\Delta t$. Then, Eq. (\ref{eq7}) is not directly solved but iterated
repeatedly to obtain the asymptotic solution $f({\bf R},t \rightarrow
\infty)$.

In the Quadratic Diffusion Monte Carlo algorithm we have used, the
Green's function $G({\bf R}^{\prime},{\bf R},\Delta t)$ is approximated
by:
\beqn
 \lefteqn{\exp \left( -A \Delta t \right) =} \\[0.6cm]
 & &   \exp \left( -A_3 \frac{\Delta
t}{2} \right ) \, \exp \left( -A_2 \frac{\Delta t}{2} \right ) \,
\exp \left( -A_1 \Delta t \right ) \,
\exp \left( -A_2 \frac{\Delta t}{2} \right ) \,
\exp \left( -A_3 \frac{\Delta t}{2} \right ) \, . \nonumber
\label{eq8}
\eeqn
This decomposition, which is not unique \cite{Chi90}, is exact up to
order $(\Delta t)^2$. Assuming (9), Eq. (\ref{eq6}) becomes:
\beqn
f({\bf R}^{\prime},t+\Delta t) & = & \int \left[
G_3 \left ({\bf R}^{\prime},{\bf R}_1,\frac{\Delta t}{2} \right )
G_2 \left ({\bf R}_1,{\bf R}_2,\frac{\Delta t}{2} \right )
G_1 \left ({\bf R}_2,{\bf R}_3,\Delta t \right ) \right .  \\[0.6cm]
 &  & \ \ \  \left .
G_2 \left ({\bf R}_3,{\bf R}_4,\frac{\Delta t}{2} \right )
G_3 \left ({\bf R}_4,{\bf R},\frac{\Delta t}{2} \right )  \right] \,
f({\bf R},t)\, d{\bf R}_1 \ldots d{\bf R}_4 d{\bf R} , \nonumber
\label{eq9}
\eeqn
with
\beqn
G_1({\bf R}^{\prime},{\bf R}, t) & = & (4 \pi D
t)^{-\frac{3N}{2}}\, \exp \left[ - \frac{({\bf R}^{\prime}-{\bf R})^2}{4 D
t} \right] ,     \\[0.6cm]
G_2({\bf R}^{\prime},{\bf R},t) & = & \delta \left(
{\bf R}^{\prime}-{\bf R}(t) \right), \ \ \ \ \ \ \mbox{where} \left \{
\begin{array}{l}
{\bf R}(0)={\bf R}  \\[0.4cm]
\frac{\displaystyle d {\bf R}(t)}{\displaystyle d t}=D\, {\bf
F}({\bf R}(t)) , \end{array}
\right.
\label{eq10}
\eeqn
and
\beq
G_3({\bf R}^{\prime},{\bf R}, t) = \exp \left[ -(E_L({\bf R})-E)\, t
\right ]\, \delta({\bf R}^{\prime}-{\bf R}) .
\label{eq11}
\eeq

In our Monte Carlo computations, $f({\bf R},t)$ is represented by
$n_w$ walkers
${\bf R}_i$ , each one representing a set of the $3N$
coordinates of the $N$ particles.
The algorithm used  for  the
implementation of Eq.(10) goes through the following steps:

\begin{enumerate}

\item{ Move the walkers, under the drift
force ${\bf F}({\bf R})$, during an interval $\Delta t/2$
with accuracy $(\Delta t)^2$.}

\item{ Apply to each walker a displacement $\chi$ randomly drawn from
the $3N$ Gaussian distribution $\exp (-\chi^2 /(4D\Delta t))$.}

\item{ Repeat step 1.}

\item{ Randomly replicate each  walker  $n_r$  times,  in such a way
that
\begin{displaymath}
\langle n_r \rangle = \exp \left[ -\Delta t\, \left(
\frac{E_L({\bf R}^{\prime})+E_L({\bf R})}{2} -E \right) \right] \, .
\end{displaymath}
}

\item{ Go to step 1 for the next walker ${\bf R}_i$ , until the set of
walkers is
exhausted. The new set obtained corresponds to $f({\bf R},t+\Delta t)$.}
\end{enumerate}

The whole procedure is repeated as many times as  it is needed  to
reach the
asymptotic limit ($t \rightarrow \infty$). From then on, the walkers
${\bf R}_i$  are
used to obtain the expectation values of the magnitudes to be
determined.

In order to establish the preciseness of the method several aspects have
to be considered. First, Monte Carlo information about $f({\bf R},t)$
only allows measurements of quantities by means of mixed estimators,
i.e., $\langle \psi |A|\Psi \rangle$. Thereby, only when the operator
$A$ is the Hamiltonian the mixed estimator gives the exact expectation
value. To obtain other ground state properties a simple linear
extrapolation \cite{Ce79} has been currently used:
\beq
\langle \Psi | A | \Psi \rangle = 2\, \langle \psi | A | \Psi \rangle -
\langle \psi | A | \psi \rangle
\label{eq12}
\eeq
This method involves the performance of a Variational Monte Carlo (VMC)
calculation to determine the variational expectation value $\langle \psi
| A | \psi \rangle $. It is interesting to notice that a VMC calculation
can be carried out with the same algorithm described for DMC only
suppressing the branching term (\ref{eq11}).

In Table I, the results for the potential, kinetic and total energies
per particle obtained with VMC and QDMC methods are shown together with
a GFMC result \cite{Ka81}. They correspond to an Aziz potential
calculation with $N=128$ at density
$\rho=0.365\ \sigma^{-3}$ ($ \sigma=2.556$ \AA). The
trial wave functions $\psi_{J1}$ and $\psi_{J2}$ contain different
two-body correlation factors, and $\psi_{JT}$ includes also three-body
correlations. Explicit expressions of these trial functions,
together with the values of the parameters involved in
them, are given in Appendix B.

As it is shown in Table I, there are not significative discrepancies
between the QDMC results for the total energy.
The perfect  agreement between  the QDMC results and the
GFMC value is also remarkable.
Equation (\ref{eq12}) is used to estimate the
kinetic and potential contributions to the total energy. In spite ot its
simplicity, this method gives very
similar values for the partial energies even when
trial wave functions as different as the ones reported
in Table I are used as importance sampling.
New methods to avoid the slight
influence of the trial wave function in the extrapolated estimators
have been recently suggested by Barnett {\em et al.} \cite{BRL91} and
Zhang and Kalos \cite{LK93}.

The effect of a finite volume simulation box has also been considered,
raising the number of particles $N$ from $N=128$ (which has been used
for the bulk of the calculation) up to $N=190$ for $\psi_{J2}$ and
$\rho=0.365\ \sigma^{-3}$. The differences
encountered were compatible with the size of the statistical
fluctuations reported in Table I.

Another important parameter in the calculation is the population of
walkers $n_w$. All the results reported in the present work have been
obtained after a preliminary analysis of the influence of the
population in the average energy of the system, and the final results
correpond to the empirical asymptotic value of $n_w$. This asymptotic
population decreases when one improves the quality of the trial wave
function $\psi$. In fact, whereas $n_w=400$ for $\psi_{J1}$ and
$\psi_{J2}$ this value is reduced to $n_w=250$ in the $\psi_{JT}$ case.
Actually, we have developed a parallel QDMC code based on the equal role
played by the different walkers. The calculation has been carried out on
a massively parallel computer CM2, which best
performance is obtained when a large number of walkers
is considered.
The length of the series have been 12000-15000
for $\psi_{J1}$ and $\psi_{J2}$ and 8000-10000 for $\psi_{JT}$.

A final but not less important point is the time-step dependence of the
QDMC algorithm. In Fig. 1 it is shown a characteristic result of the
total energy as a function of the time-step $\Delta t$. The time $t$ is
measured in reduced units $\tau$, where
\beq
\tau=\frac{m \, \sigma^2}{2 \, \hbar} \, .
\label{eq13}
\eeq
As one can see in the Figure, there is a clear
departure from the linear time dependence supplied by the linear DMC
algorithms. If a second order polynomial fit $E/N=(E/N)_0+A
(\Delta t)^2$
(solid line in the Figure) to the QDMC results is performed, one obtains
an extrapolated value of $(E/N)_0=(-7.124 \pm 0.003)\ K$, which is
indistinguishable from the values obtained working with $\Delta
t=(1-2)\cdot 10^{-3}\ \tau$. Therefore, it is plausible to calculate
the properties
of the system accurately using a single value for $\Delta t$, lying
in the
stated range, without the necessity of a complete analysis in time to
extrapolate the correct results.

\section{Results}

In this section the numerical results for the energy and for the
structure properties using the Aziz and Aziz II potentials are
presented. First, we analyse the differences between the two interatomic
potentials and then, the contribution to the total energy of several
models for the three-body interactions. In all the calculations reported
below the trial function $\psi_{J2}$ (see Appendix B) has been used as
importance sampling. The average population size ($n_w$) ranges from
400, near the equilibrium density ($\rho=0.365\ {\sigma}^{-3}$),
to 900 for the highest densities. On
the other hand, the value for the time-step has been taken as $\Delta t
=1.25 \cdot 10^{-3}\ \tau$ around the equilibrium density and $\Delta
t=1.0 \cdot 10^{-3}\ \tau$ for higher densities. No significant
deviations in the results of  the energy are observed when these $\Delta
t$ values are doubled.

\begin{center}
{\large {\bf
               A. Two-body potentials: Aziz vs. Aziz II}}
\end{center}

As it has been commented in the Introduction, the differences between
the Aziz and Aziz II potentials are not very large. However, slight
differences in the values of the parameters entering into $V(r)$
produce
relatively large changes in the energy as it was asserted by Kalos
{\it et al.} \cite{Ka81} in their GFMC calculation of the equation of
state of liquid $^4$He using the Aziz potential. The energies obtained
for both potentials, together with the experimental results of Ref.
\cite{OY} are reported in Table II. In parenthesis there are the GFMC
results \cite{Ka81} for the Aziz potential. The GFMC and QDMC
calculations are in good agreement, but a small deviation between both
results is observed at high densities. The kinetic and potential
energies are also given in the Table. The potential energy has been
calculated by means of the extrapolated estimator (Eq. \ref{eq12}), and
the kinetic energy comes from the difference between the total and the
potential energies. A comparison between the partial
energies of  the two potentials reveals that, while the kinetic energy
is practically the same, the Aziz II potential energy is, in absolute
value, larger than the Aziz one. In particular, the Aziz II potential
lowers the potential energy with respect to the Aziz case in a quantity
which
grows from $\sim 0.19\ K$ at $\rho=0.365\ \sigma^{-3}$ to $\sim 0.23\ K$
at the highest density $\rho=0.490\ \sigma^{-3}$.
The partial energies for both potentials
satisfy the lower bound for the kinetic energy
and the upper bound for the potential energy
($T/N \geq 13.4\ K$ and $V/N \leq -20.6\ K$, at the equilibrium density)
\cite{BFP}.

Concerning the total energies, one can observe that the experimental
values are approximately located at the middle of the Aziz and Aziz II
results. This fact is clear from Fig. 2 where the equation of state of
liquid $^4$He is shown in comparison with the experimental results. The
lines in the Figure correspond to numerical fits to the results reported
in Table II, excluding the highest density (0.490 $\sigma^{-3}$) because
it is quite far from the experimental freezing density $\rho_l=0.430\
\sigma^{-3}$. In the majority of microscopic calculations on liquid He
it has been used a polynomial fit of the form
\beq
  e=e_0+B \, \left( \frac{\rho-\rho_0}{\rho_0} \right) ^2 +C \, \left(
\frac{\rho-\rho_0}{\rho_0} \right) ^3 \, ,
\label{eq16}
\eeq
where $e=E/N$ and $\rho_0$ is the equilibrium density, to determine the
equation of state. On the other hand, in calculations based on Density
Functional Theory the form
\beq
e= b \, \rho + c\, \rho^{1+\gamma}  \, ,
\label{eq17}
\eeq
proposed by Stringari and Treiner \cite{ST}, has proved to be very
efficient in describing properties of homogeneous and inhomogeneous
(including an additional surface term in Eq. \ref{eq17}) liquid $^4$He.
In the case of the Aziz II potential both analytic forms are compatible
with the results of the energy, taking into account their respective
errors. However, we have checked that only the function (\ref{eq17})
provides the correct result for the energy when densities
lower
than 0.328 $\sigma^{-3}$ are considered. Therefore, all the results
presented below, concerning the equation of state, are derived
starting on the second option (Eq. \ref{eq17}).

The values of the parameters which best fit our Aziz II potential
results are:
\beqn
 b & = & (-27.258 \pm 0.017 )\ K \sigma^3     \nonumber \\[0.4cm]
 c & = & (114.95 \pm 0.22 ) \ K \sigma^{3(1+\gamma)} \nonumber
\\[0.4cm]
 \gamma & = & 2.7324 \pm 0.0020  \, .
\label{eq18}
\eeqn

The same analysis has been performed by taking the energy results of
the
Aziz potential. In this case, neither the polynomial form (\ref{eq16})
nor the Stringari's (\ref{eq17}) are statistically compatible with
our results. This fact is clearly reflected in Fig. 2, where several
Aziz points (the size of each point is larger than its error bar) are
not intersected by the result of the fit (represented with a dashed
line). In spite of this severe restriction, and to make possible the
comparison with the equation of state provided by the Aziz II potential,
the optimum values
\beqn
 b & = & (-26.947 \pm 0.016 )\ K \sigma^3     \nonumber \\[0.4cm]
 c & = & (115.72 \pm 0.21 ) \ K \sigma^{3(1+\gamma)} \nonumber
\\[0.4cm]
 \gamma & = & 2.7160 \pm 0.0020
\label{eq19}
\eeqn
are taken.

We have also fitted the same type of function to the experimental
results of Ref. \cite{OY}. In this case, the parameters $b$ and $c$ have
been fixed to reproduce the equilibrium density and the energy at this
density, whereas the parameter $\gamma$ has been obtained by means of a
numerical fit to all the energies reported in that work. The values
obtained are
\beqn
 b & = & -26.746 \ K \sigma^3     \nonumber \\[0.4cm]
 c & = & 116.69  \ K \sigma^{3(1+\gamma)} \nonumber
\\[0.4cm]
 \gamma & = & 2.7773       \, .
\label{eq20}
\eeqn

Once the equation of state $e(\rho)$ is known, it is straightforward to
calculate the isothermal compressibility, defined as
\beq
    \kappa (\rho) = \frac{1}{\rho} \, \left( \frac{ \partial
\rho}{\partial P} \right)_T    \, ,
\label{eq21}
\eeq
where $P(\rho)=\rho^2 \  (\partial e \, / \, \partial \rho)$ is the
pressure, and the velocity of sound given by
\beq
c(\rho) = \left( \frac{1}{m \, \kappa \, \rho} \right) ^ {1/2} \, .
\label{eq22}
\eeq
In Table III the results of the
pressure, the compressibility and the velocity of sound of the two Aziz
potentials are compared with the corresponding experimental values at
the experimental equilibrium density ($\rho_0^{exp}=0.3646\
\sigma^{-3}$). It is remarkable the accuracy provided by the Aziz II
potential, giving results for these quantities which
are indistinguishable from the experiment. Conversely, the equation of
state corresponding to the Aziz potential supplies results which are
slightly
worse. The differences between the equations of state for the two
potentials remain when the density increases, as one can see for
$P(\rho)$ in Fig. 3 and for $\kappa (\rho)$ in Fig. 4. The equation of
state corresponding to the Aziz II potential gives an excellent
description of these magnitudes for all the values of the density here
considered.

Apart from the ground state energy, the Monte Carlo methods yield other
interesting information. The radial distribution function
\beq
  g(r_{12})= \frac{N\,(N-1)}{\rho^2} \ \frac{ \int | \, \Psi ({\bf
r}_1,\ldots,{\bf r}_N )\, |^2 \ d{\bf r}_3 \ldots d{\bf r}_N}{ \int | \,
\Psi
({\bf r}_1,\ldots,{\bf r}_N ) \, |^2 \ d{\bf r}_1 \ldots d{\bf r}_N}
\label{eq23}
\eeq
and its Fourier transform, the static structure function
\beq
S(k)=1+\rho \, \int d{\bf r} \ e^{i {\bf k} \cdot {\bf r}} \left( g(r)
-1 \right)
\label{eq24}
\eeq
are fundamental in the study of fluids. The calculation of these
quantities is more involved than the calculation of the energy
\cite{Ce79,KLJ},
but the extrapolation procedure (Eq. \ref{eq12})  allows results which
are practically
independent of the trial function used as importance sampling.

The radial distribution
function $g(r)$, obtained in a Aziz II calculation at a density
$\rho=0.365 \ \sigma^{-3}$, is shown in Fig. 5 in comparison with an
experimental determination at $T=1.0 \ K$ by Svensson {\it et al.}
\cite{SGR}. There is a good agreement between the calculated and the
experimental $g(r)$, mainly in the first peak. In Fig. 6, the structure
function $S(k)$, obtained by means of a Fourier transform of the $g(r)$
shown in Fig. 5, is plotted together with the experimental measure of
Ref. \cite{SGR}. Due to the finite size of the simulation box, there are
not reliable results for $S(k)$ for $k \ltsim 1$ \AA$^{-1}$. The
theoretical $S(k)$ is again very close to the experimental result, but
the height of the experimental main peak is slightly higher. On the
other hand, other experimental determinations of $S(k)$ \cite{WH}
point to lower values of the intensity of the first peak, even below our
results. In fact, analysis of the influence of the temperature $T$ in
$S(k)$ \cite{WH,GR} indicate that the largest variation of the structure
function with $T$ is placed in the vicinity of the first peak.

The one-body density matrix $\rho(r)$ defined as
\beq
\rho(r_{11^{\prime}})=N\ \frac{\int \Psi({\bf r}_1^{\prime},\ldots,{\bf
r}_N)
\, \Psi({\bf r}_1,\ldots,{\bf r}_N) \ d{\bf r}_2 \ldots d{\bf r}_N}
{\int | \Psi({\bf r}_1,\ldots,{\bf r}_N)|^2 \
 \ d{\bf r}_1 \ldots d{\bf r}_N}
\label{eq25}
\eeq
and its Fourier transform, the momentum distribution
\beq
n(k)=(2 \pi)^3 \,\rho\, n_0\, \delta({\bf k}) +\rho \, \int d{\bf r} \
e^{i {\bf k} \cdot {\bf r}} \, \left( \rho(r)-\rho(\infty) \right)
\label{eq26}
\eeq
can also be computed using the configurations generated by the QDMC
code. The function $\rho(r)$ is obtained as the expectation value of the
operator
\beq
\left \langle \frac{ \Psi({\bf r}_1,\ldots,{\bf r}_i+{\bf r},\ldots,{\bf
r}_N )}{\Psi( {\bf r}_1,\ldots,{\bf r}_N ) }  \right \rangle
\label{eq27}
\eeq
evaluated on the configuration space over a set of random desplacements
of the particle $i$. The
condensate fraction $n_0$, i.e., the fraction of particles occupying the
zero momentum state, may be extracted from $\rho(r)$ by means of the
asymptotic condition
\beq
n_0= \lim_{r \rightarrow \infty} \rho(r)   \, .
\label{eq28}
\eeq

In Fig. 7 the momentum distribution obtained via the Eq. (\ref{eq26}) is
plotted, as $k\,n(k)$, for three values of the density. The correlations
between the particles make the population of states with high momenta
increase with the density. The shoulder observed at
$k \simeq 2$ \AA$^{-1}$ for the three curves, which has
been observed in other theoretical calculations of $n(k)$ \cite{PNK,PW},
has been attributed in the past to the zero-point motion of the rotons
\cite{ROT}. On the other hand, it has been proved that if the condensate
fraction is non-zero, $n(k)$ diverges as $1\,/\,k$ when $k \rightarrow
0$ \cite{NK0}. Again, the finite value of the simulation cell precludes
the possibility of reproducing this behaviour.

We have also determined the condensate fraction
from the extrapolated estimation of $\rho(r)$ and the
relation (\ref{eq28}).
At the equilibrium density, we get $n_0=0.084 \pm 0.001$ which is a
value
slightly smaller than the one obtained in a GFMC calculation ($0.092
\pm 0.001$) \cite{PW} using the Aziz potential. The discrepancy between
the two results are not due to the use of different potentials. In fact,
we have calculated $\rho(r)$ for the two Aziz potentials and no
significant differences appear. The same conclusion holds for the
radial distribution function $g(r)$.

A final point of interest is the density dependence of the condensate
fraction . In Fig. 8, the change in the value of $n_0$ is shown for
a wide range of densities. The condensate fraction decreases with the
density, following a law nearly quadratic in $\rho$. In the Figure, a
quadratic fit to the results is shown as a ``guide to the eye''.

\begin{center}
{\large {\bf
               B. Three-body interactions}}
\end{center}

The importance of three-body interactions in helium has been discussed
for  a long time. It has been argued that these interactions would be
present in He but its relative contribution to the total energy is still
open to question. The most widely known model for the three-body
potential is the triple-dipole interaction derived by Axilrod and
Teller \cite{AT} considering
perturbative theory. The Axilrod-Teller (AT) potential, which has
been usually considered as the major contribution to the energy coming
from the
three-body interactions, provides a positive correction to the potential
energy. The amount of this effect was calculated for the first time by
Murphy and Barker \cite{MB} by means of a Variational Monte Carlo
calculation. Afterwards, that contribution was estimated by Kalos {\em
et al.} \cite{Ka81,KLJ} in a Rayleigh-Schr\"{o}dinger
perturbative calculation starting on GFMC configurations. From this
analysis it was pointed out that, on the one hand, the GFMC prediction
for the expectation value of the three-body potential $\langle V_3
\rangle$ was in accordance with the Murphy and Barker's variational
results and, on the other, there were no relevant  differences
between the results coming from a Lennard-Jones and an Aziz potential
calculations. Another conclusion of these GFMC works was that the
inclusion
of three-body potential contributions on the total energy worsened the
two-body results along the whole equation of state.

In spite of the AT potential being the dominant contribution to $\langle
V_3 \rangle$, it has been proved that at short interparticle separations
a non-additive and attractive force emerges. This short-ranged
three-body interaction , usually known as exchange interaction, is due
to the influence in the charge densities of two interacting atoms by
the presence of a third near particle. Bruch and McGee \cite{BM}
proposed a model
potential (BM) to account for this effect, fitting the parameters of
the exchange part to the atomic calculations of the energy of three He
atoms at very short distances from Novaro and Beltran-Lopez \cite{NBL}.
Loubeyre \cite{Lou} has proved that the BM three-body potential, in
conjunction with the Aziz potential, accurately describes
solid helium
at high pressures and room temperature. The explicit forms for the AT
and BM potentials are given in Appendix A.

As it has been previously discussed, the Aziz II results for
the energy
per particle are below the experimental results for all the densities
considered (see Fig. 2). Therefore, the inclusion of a repulsive
contribution to the potential energy, arising from three-body
interactions, could bring the theoretical results nearer to the
experimental ones. In Table IV the results for the total ($E/N$) and
potential ($V/N$) energies are reported in comparison with the
experimental values of the energy. In all cases, the three-body
potential energy is obtained by means of a Rayleigh-Schr\"{o}dinger
perturbative calculation, following the method described by Kalos
{\em et al.} \cite{KLJ}. As one can see, the AT potential produces an
increase in the energy, leading to values which are slightly
higher than the experiment. Moreover, the difference between the Aziz
II+AT and experimental values increases appreciably with the
density, yielding to poor results for derivative magnitudes of the
energy as the pressure or the compressibility. The results of the
energy, using the BM potential, appear in the second column of Table IV.
The exchange part of the BM potential practically cancells
the repulsive
contribution of the dispersion term (AT) becoming even dominant at the
highest densities. The resulting energies lie very near to the
two-body
calculation but also in this case, as in the AT one, with a worsening
reproduction of the dependence of the energy with the density.
Therefore, neither the simple AT potential nor the more elaborated one
(BM) improve, in a significant way, the Aziz II results. In fact, it
seems more convincing that, in the density regime of liquid $^4$He, the
main three-body contribution comes from the AT potential,
the exchange part of the BM potential being too large. We should
notice that
the parameters of the BM potential have been fitted to reproduce the
energy of helium trimers with interparticle separations
considerably less
than the characteristic mean distance between the atoms in the liquid.
Then, it is uncertain that the same parameters, or
even the same analytical form, could be used to study the liquid phase.

In the third column of Table IV, labelled as MBM, we report the results
which are obtained by using the BM potential with a modified value
$A^{\prime}=A\, / \, 3$ (see Appendix A).
Now, the energy at the experimental equilibrium density
reproduces the experimental result and a quite good description is also
obtained at higher densities. In Fig. 9, the equation of state obtained
with the Aziz II+MBM model is depicted together with the experimental
and the Aziz II results. The values of the parameters of the fit for the
Aziz II+MBM calculation are:
\beqn
 b & = & (-27.202 \pm 0.017 )\ K \sigma^3     \nonumber \\[0.4cm]
 c & = & (114.11 \pm 0.21 ) \ K \sigma^{3(1+\gamma)} \nonumber
\\[0.4cm]
 \gamma & = & 2.6961 \pm 0.0020
\label{eq29}
\eeqn

The Aziz II+MBM results for the pressure and the compressibility are
plotted in Fig. 10 and Fig. 11, respectively, in comparison with the
experimental values. One can observe that there are slight differences
between the theoretical and the experimental results, which are more
evident in the pressure case. In fact, these discrepancies reflect the
departure of the Aziz II+MBM total energies from the experimental
values when
the density increases. This small effect on the energy, which can be
observed in Fig. 9, is enlarged when the derivative magnitudes of the
energy as $P(\rho)$ or $\kappa (\rho)$ are calculated.

\section{Discussion}
The properties of bulk liquid $^4$He have been investigated by means of
the Diffusion Monte Carlo (DMC) method. It has been proved that the
extension of the DMC algorithm up to second order (QDMC) allows for the
possibility of calculating the energy without the extrapolation to
$\Delta t =0$, required in the linear DMC codes. We have applied the
QDMC method in order to perform a comparative analysis between the Aziz
and the
Aziz II two-body potentials. The calculations have been extended to a
wide range of densities in order to contrast the theoretical predictions
on the equation of state provided by the two Aziz potentials.  The
results unambiguously demonstrate that the new Aziz potential gives
better results
than the old one, especially when the dependence of the pressure and the
compressibility on the density is  considered. In particular, the
Aziz II results for $P(\rho)$ and $\kappa(\rho)$ are
indistinguishable from the experimental values. However, the
results for the energy are below the experimental determinations. This
difference could suggest the presence of three-body interactions in He.

We have performed a Rayleigh-Schr\"{o}dinger perturbative estimation of
the three-body potential energy using two different models. The results
obtained have shown that neither the triple-dipole potential of
Axilrod-Teller nor the Brunch-McGee potential, which includes the
exchange interaction at short distances, improve the equation of state
given by the Aziz II potential. To make the three-body
correction compatible with the experimental results a simple change
in the parameters entering into the BM potential has been examined
(MBM potential). The Aziz II+MBM model provides a good description of
the equation of state ($E\,/\,N)(\rho)$ but the results for $P(\rho)$
and
$\kappa(\rho)$ worsen with respect to the ones calculated with the Aziz
II only. On the other hand, we would
point out that the Aziz II results for the energy are shifted with
respect to the experiment in
a constant value for all the densities. This
fact explains the excellent description of $P(\rho)$ and $\kappa
(\rho)$ given by the new Aziz potential. The conclusion is that to
account for the experimental energies, a constant value for
$\langle V_3 \rangle$  would be required, although from
the theoretical point of view it seems more plausible a correction
which becomes larger when the density increases.

Concerning  other properties as the radial distribution function or
the momentum distribution, no significant differences between the
results given by the two Aziz potentials are observed. Overall, the
agreement between the QDMC results and the experiment is quite
satisfactory.
Finally, we would remark that the accuracy of the Aziz II
potential in describing the bulk $^4$He liquid phase makes it
recommendable for future calculations of the solid phase or films,
especially when the derivative magnitudes of the energy are among
the main objectives.

\section*{Acknowledgements}

This work has been supported in part by DGICYT (Spain) Grants No.
PB92-0761, No. PB90-06131 and No. TIC-0299/1989. Most of the simulations
were
performed on a multiprocessor CM2 from Thinking Machines of the CEPBA
(Centre Europeu de Paral.lelisme de Barcelona).

\pagebreak

\section*{Appendix A: Two- and three-body potentials}

\def\theequation{\thesection.\arabic{equation}}
\def\thesection{\Alph{section}}
\setcounter{section}{1}
\setcounter{equation}{0}

The form of the HFDHE2 (Aziz) potential \cite{Az79} is
\beq
V(r)= \varepsilon \, \left[ A \, \exp (-\alpha x) -F(x) \,
\sum_{j=0}^{2}
\left( C_{2j+6}\, / \, x^{2j+6} \right) \, \right] \, ,
\label{a1}
\eeq
where
\beq
F(x) = \left\{  \begin{array}{ll}
                \exp \left[ - \left( \frac {D}{x}-1 \right)^2 \right] \,
&  x < D  \,    \\
                1    \,    &  x \ge D  \,
               \end{array}
\right.
\label{a2}
\eeq
with
\beq
    x=\frac{r}{r_m}  \, .
\label{a3}
\eeq

The values of the parameters for the Aziz potential are
\beq
\begin{array}{lcl}

 \varepsilon=10.8 \ K    &  &  C_6=1.3732412    \\[0.3cm]
 r_m=2.9673\ \mbox{\AA}  &  &  C_8=0.4253785     \\[0.3cm]
 D=1.241314              &  &  C_{10}=0.1781       \\[0.3cm]
 \alpha=13.353384        &  &  A=0.5448504 \cdot 10^6  \, .
\end{array}
\label{a4}
\eeq

The HFD-B(HE) (Aziz II) potential \cite{Az87}, which is quite similar in
form to the Aziz potential, is given by
\beq
V(r)= \varepsilon \, \left[ A \, \exp (-\alpha x+\beta x^2) -F(x) \,
\sum_{j=0}^{2}
\left( C_{2j+6}\, / \, x^{2j+6} \right) \, \right] \, ,
\label{a5}
\eeq
where the function $F(x)$ and $x$ are formally the same as in the
Aziz
potential (Eqs. \ref{a2},\ref{a3}). The values of the parameters for
the Aziz II potential are
\beq
\begin{array}{lcl}

 \varepsilon=10.948 \ K    &  &  C_6=1.36745214   \\[0.3cm]
 r_m=2.963\ \mbox{\AA}     &  &  C_8=0.42123807    \\[0.3cm]
 D=1.4826                  &  &  C_{10}=0.17473318   \\[0.3cm]
 \alpha=10.43329537        &  &  A=1.8443101 \cdot 10^5  \\[0.3cm]
 \beta=-2.27965105   \, .  &  &
\end{array}
\label{a6}
\eeq

The models for the three-body interactions we have used are those
given by the Axilrod-Teller (AT) \cite{AT} and  Brunch-McGee (BM)
\cite{BM} potentials. The form of the AT potential is
\beq
V_3(r_{12},r_{13},r_{23})= \frac{ \nu \, \left( 1+3 \, \cos \phi_1 \cos
\phi_2 \cos \phi_3 \right)}{r_{12}^3 \, r_{13}^3 \, r_{23}^3} \,
\label{a7}
\eeq
where $\phi_1$, $\phi_2$ and $\phi_3$ are the interior angles of the
triangle formed by the three atoms. We use the Leonard's
helium value $\nu=0.327\ K \sigma^9$ \cite{Leo}, assuming the
radial distances $r_{ij}$ in $\sigma$ unities.

The BM potential is given by
\beqn
V_3(r_{12},r_{13},r_{23}) & = & \left[ \frac{\nu}{r_{12}^3 \,
r_{13}^3
\, r_{23}^3} - A \, \exp \left( - \alpha \left( r_{12}+r_{13}+r_{23}
\right) \, \right) \, \right]  \hspace{2.5cm}  \nonumber \\[0.4cm]
  &  & \hspace{2.5cm} \times \, \left( 1+3 \, \cos \phi_1 \cos \phi_2
\cos \phi_3 \right) \, .
\label{a8}
\eeqn
 The values for the two new parameters appearing in the BM potential are
\beq
\begin{array}{lcl}
 A = 9676545.53 \ K    &  &  \alpha = 4.948 \ \sigma^{-1}   \, .
\end{array}
\label{a9}
\eeq

\section*{Appendix B: Trial functions}

\def\theequation{\thesection.\arabic{equation}}
\def\thesection{\Alph{section}}
\setcounter{section}{2}
\setcounter{equation}{0}

In this Appendix we give the explicit forms of the trial functions used
as importance sampling in the QDMC calculations as well as the values of
the parameters involved. The first one is the well
known McMillan two-body trial function \cite{McM}
\beq
  \psi_{J1} = \prod _{i<j} \exp \left[ -\frac{1}{2} \, \left(
\frac{b}{r_{ij}} \right)^5 \, \right]  \, .
\label{b1}
\eeq
We have taken the value $b=1.20 \ \sigma$ which optimizes the VMC
energy at the experimental equilibrium density.

Most of the present work has been carried out using the Reatto
two-body function \cite{Reat}
\beq
\psi_{J2}= \prod_{i<j} \exp \left[ -\frac{1}{2} \, \left(
\frac{b}{r_{ij}} \right)^5 -
\frac{L}{2} \, \exp \left( - \left( \frac{r_{ij}-\lambda}{\Lambda}
\right)^2 \right) \, \right]     \, ,
\label{b2}
\eeq
with $L=0.2$, $\lambda=2.0 \ \sigma$, $\Lambda=0.6 \ \sigma$ and $b=1.20
\ \sigma$. These values, optimal at the experimental equilibrium
density, have also been used for the other densities.

The third trial function, which was proposed by Schmidt {\em et
al.} \cite{Schm}, contains two- and three-body correlations. It is
explicitely given by
\beq
 \psi_{JT}= \psi_{J1} \, \exp \left( -\frac{1}{4} \, \lambda \sum_{k}
{\bf G}_k \cdot {\bf G}_k + \frac{1}{2} \, \lambda \sum_{i<j}
\xi^2(r_{ij}) r_{ij}^2 \, \right) \,    ,
\label{b3}
\eeq
where
\beq
{\bf G}_k= \sum_{l \neq k} \xi(r_{kl}) \, {\bf r}_{kl}
\label{b4}
\eeq
and
\beq
 \xi(r)= \exp \left[ - \left( \frac{r-r_t}{r_{\omega}} \right)^2 \,
\right]  \, .
\label{b5}
\eeq
The values for the triplet parameters, roughly optimal at the
equilibrium density, are:   $\lambda=-1.08\
\sigma^{-2}$, $r_t=0.82 \ \sigma$ and $r_{\omega}=0.50 \ \sigma$.

\pagebreak

\pagebreak

\section*{Table Captions}

{\bf Table I}: Results for the total, kinetic and potential energies
for different trial wave functions. The forms of $\psi_{J1}$,
$\psi_{J2}$ and $\psi_{JT}$ as well as the values of the parameters
entering into them are noted explicitely in Appendix B. In the last
row,
the GFMC results from Ref. \cite{Ka81} (a) and Ref. \cite{PW} (b) are
also reported. All energies are in kelvin per particle.

\vspace{0.2cm}

\noindent
{\bf Table II}: Results for the total and partial energies from
the QDMC calculations with the Aziz potential, the Aziz II potential and
experiment \cite{OY}. The numbers quoted in parenthesis are taken from
Ref. \cite{Ka81}. All energies are in  kelvin per particle.

\vspace{0.2cm}

\noindent
{\bf Table III}: QDMC results for the pressure $P$, the compressibility
$\kappa$ and the velocity of sound $c$ at the experimental equilibrium
density using the Aziz and Aziz II potentials. The last row contains the
experimental values derived from the experimental equation of state
(20).

\vspace{0.2cm}

\noindent
{\bf Table IV}: Energies from the QDMC calculations with the Aziz II
potential including the perturbative estimation of the expectation value
of several models for the three-body interactions (AT, BM, MBM; see
text).The last column contains the experimental values. All energies are
in kelvin per particle.

\pagebreak

\section*{Figure Captions}

{\bf Fig. 1}. Time-step dependence in the QDMC method. The solid line is
a second order polynomial fit to the calculated points.

\vspace{0.2cm}

\noindent
{\bf Fig. 2}. Equation of state for liquid $^4$He. The circles are the
QDMC results with the Aziz potential and the dashed line is a fit to the
calculated energies. The solid circles correspond to the QDMC energies
with the Aziz II potential; the solid line is a fit to these energies.
These fits have been performed with Eq. (17),
the parameters being those given in (18) and (19) for the Aziz II and
Aziz
potentials, respectively. The experimental values, represented by solid
triangles, are taken from Ref. \cite{OY}. The error bars of the QDMC
results are smaller than the size of the symbols.

\vspace{0.2cm}

\noindent
{\bf Fig. 3}. Pressure of liquid $^4$He as a function of the density.
The circles and solid circles correspond to QDMC calculations with the
Aziz and Aziz II potentials, respectively. The dashed and solid lines
are numerical fits to the Aziz and Aziz II pressures, respectively.
The experimental results, from the experimental equation of state (20),
are represented by diamonds which are practically hidden below the Aziz
II values. The error bars of the QDMC results are smaller than the size
of the symbols.

\vspace{0.2cm}

\noindent
{\bf Fig. 4}. Isothermal compressibility of liquid $^4$He as a function
of the density. The same notation as in Fig. 3.

\vspace{0.2cm}

\noindent
{\bf Fig. 5}. Two-body radial distribution function at the experimental
equilibrium density. The solid line is the QDMC result and the solid
circles correspond to the neutron diffraction experimental determination
from Ref. \cite{SGR}.

\vspace{0.2cm}

\noindent
{\bf Fig. 6}. Static structure function at the experimental equilibrium
density. The solid line is the QDMC result, obtained by a Fourier
transform of the radial distribution function showed in Fig. 5. The
solid circles are the experimental determination from Ref. \cite{SGR}.

\vspace{0.2cm}

\noindent
{\bf Fig. 7}. Dependence  of the calculated momentum
distribution on density. The long-dashed, solid and
short-dashed lines stand for the results at densities of 0.328
$\sigma^{-3}$, 0.365 $\sigma^{-3}$ and 0.401 $\sigma^{-3}$,
respectively.

\vspace{0.2cm}

\noindent
{\bf Fig. 8}. Condensate fraction in liquid $^4$He as a function of
density. The solid line is a second order polynomial fit to the
calculated values. The error bars of the results are smaller than the
size of the symbols.

\vspace{0.2cm}

\noindent
{\bf Fig. 9}. Equation of state of liquid $^4$He. The circles are the
QDMC results with the Aziz II+MBM potentials; the solid line is a fit to
these energies. The solid circles and the dashed line correspond to the
calculation with the Aziz II potential. The solid triangles are the
experimental values form Ref. \cite{OY}. The error bars are smaller than
the size of the symbols.

\vspace{0.2cm}

\noindent
{\bf Fig. 10}. Pressure of liquid $^4$He as a function of the density.
The solid circles are the Aziz II+MBM results and the diamonds are the
experimental values from the equation of state (20). The solid line is a
numerical fit to the data. The error bars are smaller than the size of
the symbols.

\vspace{0.2cm}

\noindent
{\bf Fig. 11}. Isothermal compressibility of liquid $^4$He as a function
of the density. The same notation as in Fig. 10.

\pagebreak

\begin{center}

\begin{tabular}{@{\extracolsep{0.3cm}}|l|ccc|}    \hline \hline
{}  & $E/N$ &  $T/N$  &  $V/N$   \\[0.2cm] \hline
VMC-$\psi_{J1}$  &  -5.683$\pm$0.014  &  15.119$\pm$0.005  &
-20.802$\pm$0.009 \\[0.4cm]
VMC-$\psi_{J2}$  &  -5.881$\pm$0.005  &  15.248$\pm$0.004  &
-21.129$\pm$0.007 \\[0.4cm]
VMC-$\psi_{JT}$  &  -6.617$\pm$0.007  &  14.552$\pm$0.030  &
-21.169$\pm$0.018 \\[0.8cm]
QDMC-$\psi_{J1}$  &  -7.115$\pm$0.010  &  14.589$\pm$0.020  &
-21.704$\pm$0.020 \\[0.4cm]
QDMC-$\psi_{J2}$  &  -7.121$\pm$0.010  &  14.576$\pm$0.025  &
-21.697$\pm$0.023 \\[0.4cm]
QDMC-$\psi_{JT}$  &  -7.125$\pm$0.005  &  14.417$\pm$0.030  &
-21.542$\pm$0.020 \\[0.8cm]
GFMC  &  \ \ \ -7.120$\pm$0.024$^{(a)}$  &  14.47$\pm$0.09$^{(b)}$  &
-21.59$\pm$0.09$^{(b)}$  \\   \hline   \hline
\end{tabular}

\vspace{2cm}

{\bf Table I}

\end{center}

\pagebreak

\begin{center}

{\small

\begin{tabular}{@{\extracolsep{-0.15cm}}|c|ccc|ccc|c|}    \hline \hline
{}  & {}  & Aziz & {} &  {}  & Aziz II &  {}  &  Exp  \\[0.2cm]
$\rho(\sigma^{-3})$ & $E/N$  & $T/N$ &  $V/N$  &  $E/N$  & $T/N$  &
$V/N$ & $E/N$   \\[0.2cm]
\hline
0.328  &  -6.988$\pm$0.013  &   12.107$\pm$0.018  &
          -19.095$\pm$0.013
       &  -7.150$\pm$0.010   &   12.152$\pm$0.032  &
          -19.302$\pm$0.030
       &    {}
\\
{}  &  (-7.034$\pm$0.037) & {} &  {}  &  {}  &  {}  &  {} & {}
\\[0.3cm]
0.365  &  -7.121$\pm$0.010  &   14.576$\pm$0.025  &
          -21.697$\pm$0.023
       &  -7.267$\pm$0.013   &   14.622$\pm$0.027  &
          -21.889$\pm$0.024
       &   -7.17
 \\
{}  &  (-7.120$\pm$0.024) & {} &  {}  &  {}  &  {}  &  {} & {}
\\[0.3cm]
0.401  &  -6.892$\pm$0.013  &   17.262$\pm$0.030  &
          -24.154$\pm$0.027
       &  -7.150$\pm$0.016   &   17.302$\pm$0.038  &
          -24.452$\pm$0.035
       &  -7.03
\\
{}  &  (-6.894$\pm$0.048) & {} &  {}  &  {}  &  {}  &  {} & {}
\\[0.3cm]
0.424  &  -6.696$\pm$0.024  &   19.152$\pm$0.042  &
          -25.848$\pm$0.035
       &  -6.877$\pm$0.022   &  19.218$\pm$0.037   &
          -26.095$\pm$0.030
       &  -6.77
\\
{}  &  {} & {} &  {}  &  {}  &  {}  &  {} & {}
\\[0.3cm]
0.438  &  -6.422$\pm$0.020  &   20.447$\pm$0.036  &
          -26.869$\pm$0.030
       &  -6.660$\pm$0.017   &   20.398$\pm$0.034  &
          -27.058$\pm$0.030
       &  -6.55
\\
{}  &  (-6.564$\pm$0.058) & {} &  {}  &  {}  &  {}  &  {} & {}
\\[0.3cm]
0.490  &  -5.010$\pm$0.025  &   25.402$\pm$0.047  &
          -30.412$\pm$0.040
       &  -5.222$\pm$0.025   &   25.404$\pm$0.050 &
          -30.626$\pm$0.043
       &   {}
\\
{}  &  (-5.175$\pm$0.101) & {} &  {}  &  {}  &  {}  &  {} & {}  \\
\hline   \hline
\end{tabular}
}

\vspace{2cm}

{\bf Table II}

\end{center}

\pagebreak

\begin{center}

\begin{tabular}{@{\extracolsep{0.3cm}}|l|ccc|}    \hline \hline
{}  & P(atm) &  $\kappa$(atm$^{-1}$)  &  c(m/s)   \\[0.2cm] \hline
Aziz  &  0.878$\pm$0.073  &  0.01199$\pm$0.00004  &
241.53$\pm$0.44 \\[0.6cm]
Aziz II  &  -0.019$\pm$0.075  &  0.01241$\pm$0.00004  &
237.40$\pm$0.46 \\[0.6cm]
Exp  &  0. &  0.0124  &  237.2     \\  \hline  \hline
\end{tabular}

\vspace{2cm}

{\bf Table III}

\end{center}

\pagebreak

\begin{center}

\begin{tabular}{@{\extracolsep{0.15cm}}|c|cc|cc|cc|c|}    \hline \hline
{}  & \multicolumn{2}{c|}{Aziz II + AT} &
\multicolumn{2}{c|}{Aziz II + BM} & \multicolumn{2}{c|}{Aziz II + MBM}
& Exp \\[0.2cm]
$\rho(\sigma^{-3})$ & $E/N$  & $V/N$ &  $E/N$  &  $V/N$  & $E/N$  &
$V/N$ & $E/N$   \\[0.2cm]
\hline
0.328  &  -7.045  &  -19.197
       &  -7.122  &  -19.274
       &  -7.071  &  -19.223
       &    {}
\\[0.3cm]
0.365  &  -7.127  &  -21.749
       &  -7.249  &  -21.871
       &  -7.168  &  -21.790
       &  -7.17
\\[0.3cm]
0.401  &  -6.971  &  -24.273
       &  -7.141  &  -24.443
       &  -7.027  &  -24.329
       &  -7.03
\\[0.3cm]
0.424  &  -6.668  &  -25.886
       &  -6.886  &  -26.104
       &  -6.741  &  -25.959
       &  -6.77
\\[0.3cm]
0.438  &  -6.435  &  -26.833
       &  -6.675  &  -27.073
       &  -6.515  &  -26.913
       &  -6.55
\\[0.3cm]
0.490  &  -4.913  &  -30.317
       &  -5.323  &  -30.727
       &  -5.050  &  -30.454
       &   {}
\\  \hline   \hline
\end{tabular}

\vspace{2cm}

{\bf Table IV}

\end{center}


\end{document}